\documentclass[12pt]{article}
\usepackage{amsfonts}
\usepackage{amsmath}
\usepackage{amssymb}
\usepackage{cite}
\usepackage{graphicx}
\usepackage{color}
\usepackage{enumitem}
\newcommand{\me}{\mathbb{E}}
\newcommand{\var}{\mathrm{var}}

\date{July 22, 2013}

\begin{document}
\title{Can we define a best estimator\\ in simple 1-D cases ?}

\author{\'{E}ric~Lantz\thanks{E. Lantz is with Department of Optics,
  Femto-ST,
  University of Franche-Comt\'{e}/CNRS,
UFR Sciences - Route de Gray,
25030 Besan\c{c}on Cedex - France (Email: \texttt{eric.lantz@univ-fcomte.fr}).}
~and~Fran\c{c}ois~Vernotte\thanks{F. Vernotte is with UTINAM,
  Observatory THETA of Franche-Comt\'{e},
  University of Franche-Comt\'{e}/CNRS,
41 bis avenue de l'observatoire - B.P. 1615,
25010 Besan\c{c}on Cedex - France.}}

\maketitle

\section{Scope}
What is the best estimator for assessing a parameter of a probability distribution from a small number of measurements? Is the same answer valid for a location parameter like the mean as for a scale parameter like the variance? It is sometimes argued that it is better to use a biased estimator with low dispersion than an unbiased estimator with a higher dispersion. In which cases is this assertion correct ? In order to answer these questions, we will compare on a simple example the determination of a location parameter and a scale parameter with three ``optimal'' estimators: the minimum-variance unbiased estimator, the minimum square error estimator and the \textit{a posteriori} mean.

\section{Relevance}
Nowadays, it seems that processing a huge amount of data is a very common task. However, in some cases it is of great importance of being able to assess the statistical parameters of a process from a very small number of measurements.  This can occur for instance in the analysis of the very long term behavior of time series  (e.g. amplitude estimation of very low frequencies, time keeping, etc.). This paper focuses on the choice of the best estimator to be used over, say, less than $10$ measurements.

\section{Prerequisites}
The reader is expected to have a basic understanding of data statistical processing, such as the one developed in \cite{Saporta2011}.

\section{Problem statement}
\subsection{Two examples of measurement}
We consider in this note the simplest archetypal measurement situation: an unknown quantity $\mu$ is measured $N$ times, giving $N$ measurements forming a set $\{d_i,i=1\ldots N\}$, noted $\{d_i\}$ in the following. Each measurement may be written as
\begin{equation}
d_i=\alpha \mu+n_i
\end{equation}\\
where
\begin{itemize}
	\item $\alpha \mu$ is the deterministic part: $\alpha$ is  a known constant\footnote{For rendering this example relevant, we ought to keep in mind that $\alpha$ can be frequency dependent. For instance, $\alpha$ may be a transfer function $H(f)$ of a measurement apparatus for a given frequency $f$, if $\theta$ is the Fourier transform for this frequency of a time varying signal. }, $|\alpha|\leq 1$, and $\mu$ a parameter we want to estimate;
	\item $n_i$ is the random part which is supposed to be a zero-mean additive white Gaussian noise, of unknown variance $\sigma^2$, independent from one measurement to another.
\end{itemize}
In the following, we discuss successively  the estimation of the two unknown parameters appearing in this measurement process. The generic name of the unknown parameter will be $\theta$, corresponding respectively to:

\begin{itemize}
	\item \textbf{(Example 1)} $\theta=\mu$. In this first example, $\theta$ can be either positive or negative and is called a location parameter, since the probability density of the data $d_i$ can be expressed as a function of the difference of location $|d_i-\alpha\theta|$.
	\item \textbf{(Example 2)}  $\theta=\sigma^2$. This variance is positive and is called a scale parameter since it determines the precision scale relevant for the measurement of the  location parameter of the measurement process.
\end{itemize}
Before going further to the definition of estimators, let us recall the heuristic concepts of \textbf {location parameter} versus \textbf{scale parameter} and \textbf{model world} versus \textbf{measurement world}.

\subsection{Location parameter and scale parameter}
Let us consider a random variable $d$ which depends on a parameter $\theta$. Let us denote $p_\theta(d)$ its probability density function (PDF).

\subsubsection{Location parameter}
A location parameter is a parameter whose variation induces a shift of the PDF of a random variable which depends of it. It is an additive parameter.

Let us denote $p_0(d)$ the PDF obtained for $\theta=0$:
\begin{equation}
p_\theta(d)=p_0(d-\theta).
\end{equation}\\
The mean and the median of a normal distribution are location parameters.

\subsubsection{Scale parameter}
A scale parameter is a positive parameter which controls the flattening or the narrowness of a PDF, for example the variance of a normal distribution. It is a multiplicative parameter.
Let us denote $p_1(d)$ the PDF of the estimator of this parameter, obtained for $\theta=1$. We have :
\begin{equation}
p_\theta(d)=\frac{1}{\theta}p_1\left(\frac{d}{\theta}\right).
\label{scale}
\end{equation}\\
$p_1(d)$ is for example a $\chi^2$ distribution if $d$ is the estimator of the variance of a normal distribution.

\subsection{Model world and measurement world}
Two problems are generally addressed: the \textit{direct problem}, which aims to forecast the measurement data knowing the parameter and the \textit{inverse problem}, which aims to estimate the parameter knowing the measurement data. In the same vein, Tarantola distinguishes the \textit{model space}, i.e. the space in which the parameter is given, from the \textit{data space}, i.e. the space in which the measurement data are given \cite{Tarantola1987}. In the following, we will use the terms ``model world'' and ``measurement world''.

\subsubsection{Model world (direct problem)}
In the model world, the question is \emph{``Knowing the parameter $\theta$, how are the measurements $\{d_i\}$ distributed?''}.
We have to define the conditional PDF: $p(d_i|\theta)$ where the vertical bar means ``knowing'' (i.e. the probability of obtaining the measurement $d_i$ knowing that the parameter is equal to $\theta$).

In the model world, the model parameter $\theta$ is considered as a definite quantity whereas the measurements $\{d_i\}$ are realizations of a random variable $d$.

However, the parameter $\theta$ is precisely the unknown quantity that we want to estimate. Supposing that this parameter is known has sense only in theory and simulations.

\subsubsection{Measurement world (inverse problem)}
In the measurement world, the question is \emph{``Knowing the measurements $\{d_i\}$, how to estimate a confidence interval over $\theta$?''}.
We need thus to reverse the previous conditional PDF for defining $p(\theta|\{d_i\})$ which describes the probability that the parameter is equal to $\theta$ knowing that the measurements are $\{d_i\}$.

This is the right question of the metrologist!

Let us notice that in the measurement world, the parameter $\theta$ is considered as a random variable whereas the measurements $\{d_i\}$ are data, i.e. totally determined values.

\subsection{Three ``optimal'' estimators\label{sec:def}}
We want to construct an ``optimal'' estimator $\hat{\theta}$ as a function of the measurements: $\hat{\theta}=f(\{d_i\})$ and we will see rapidly  that the usual optimality criteria do not work equally on both examples. Three estimators are often used as optimal, even if it is well known that they are generally different from each other for small $N$. Let us first see the main properties of these three estimators. Their mathematical calculations for both examples will be described in Section \ref{sec:app3est2ex}.

\begin{itemize}
	\item \textbf{(Estimator 1) minimum-variance unbiased estimator}.

Properties:
	\begin{enumerate}[label=\textit{P1.\arabic{enumi})} , leftmargin=1.5cm]
		\item the estimator is unbiased: $\me(\hat{\theta})=\theta$, where $\me$ stands for mathematical expectation.
		\item among the unbiased estimators, it has the smallest variance: $\me\left[\left(\hat{\theta}-\me(\hat{\theta})\right)^2\right]$ minimum.
	\end{enumerate}

Since we consider the mathematical expectation of $\hat{\theta}$, it means that we consider this estimator as a random variable, like the measurements, and thus we define these properties in the model world.

	\item \textbf{(Estimator 2) minimum mean square error estimator (MMSE)}.
A MMSE estimator is an estimation method which minimizes, in the  model world, the mean square error of the estimator regardless of a possible bias \cite{Johnson2004}.

Properties:
	\begin{enumerate}[label=\textit{P2.\arabic{enumi})} , leftmargin=1.5cm]
		\item The mean square error (MSE) $\me\left[(\hat{\theta}-\theta)^2\right]$ is minimum
		\item this estimator can be biased. Since
\begin{equation}
\me\left[(\hat{\theta}-\theta)^2\right]=\left[\me(\hat{\theta})-\theta\right]^2+\me\left[\left(\hat{\theta}-\me(\hat{\theta})\right)^2\right], \label{eq:1}
\end{equation}\\
	\end{enumerate}
the idea is to admit some bias (first term of the sum) in order to strongly diminish the variance of the estimator (second term).


	\item \textbf{(Estimator 3) \textit{a posteriori} mean}. In this so-called Bayesian approach, the measurements, and therefore $\hat{\theta}$, are no more considered as random variables (as they are in the model world), 
but as a particular realization of these random variables, i.e. known data having given values in the measurement world. 
In this measurement world, $\theta$ appears as a random variable and we aim to construct a probability law on $\theta$ with density $p(\theta)$ that takes into account these measurements: $p(\theta|\{d_i\})$ and, if available, all the information that was known before the measurements: $\pi(\theta)$. This information $\pi(\theta)$ is called ``\textit{a priori}'' and $p(\theta)=\pi(\theta)p(\theta|\{d_i\})$ is called ``\textit{a posteriori} probability density''  (i.e. after the measurement process). The \textit{a posteriori} mean is $\hat{\theta}=\me(\theta)=\int_{-\infty}^{+\infty}\theta p(\theta) d\theta$.

Properties:
\begin{enumerate}[label=\textit{P3.\arabic{enumi})} , leftmargin=1.5cm]
	\item this estimator minimizes the \textit{a posteriori} mean square error: $\hat{\theta}$ is now a constant and
\begin{equation}\\
\me\left[(\hat{\theta}-\theta)^2\right]=\left[\me(\theta)-\hat{\theta}\right]^2+\me\left[\left(\theta-\me(\theta)\right)^2\right], \label{posterior}
\end{equation}\\
	 since the variance of $\theta$ (second term of Eq. (\ref{posterior})) does not depend on the estimator, the mean square error is minimized if the first term vanishes.
\end{enumerate}
\end{itemize}

\subsection{Applying the three estimators to the two measurement processes\label{sec:app3est2ex}}
Let us show some significant differences in the use of these estimators on the two above examples.

	\subsubsection {Minimum-variance unbiased estimator on Example 1}
It is evidently:
\begin{equation}
\hat{\theta}=\frac{\bar{d}}{\alpha}=\theta+\frac{1}{N\alpha}\sum_{i=1}^N n_i,
\end{equation}\\
  where $\bar{d}$ is the sample mean, i.e. the average of the $N$ measurements.

However, this estimator cannot be employed if $N|\alpha|^2\ll 1$: though the noise term has a zero mean, its variance $\frac{\sigma^2}{N|\alpha|^2}$ is high and the error, despite its null expectation, can be high. Therefore, estimators 2 or 3 must be used. 

	\subsubsection{MMSE estimator or \textit{a posteriori} mean on Example 1}
We find (see Annex 1):
\begin{equation}
\hat{\theta}=\frac{1}{\alpha} \frac{\bar{d}}{1+\displaystyle\frac{\me(n^2)}{N|\alpha|^2\me(\theta^2)}}.\label{eq:2}
\end{equation}\\
This formula tells us that we may restore $\theta$, the true value of the signal before measurement, if the signal-to-noise ratio after measurement is high. It is known as Wiener filtering and is based on an estimation, even rough, of this signal-to-noise ratio. This kind of information does not come directly from the measurements: at a specific frequency, it is not straightforward to distinguish between the signal and the noise. It is called ``\textit{a priori}'' information (before the measurements). To simplify, we have supposed $\me_\mathrm{prior}(\theta)=0$ and the derivation of the \textit{a posteriori} mean \cite{Johnson2004} assumes Gaussian \textit{a priori} laws for $n$ and $\theta$.

Of course, if we have absolutely no information about the signal-to-noise ratio, we should consider all the output signal as carrying information and the unbiased estimator is the best. This situation rarely occurs in practice: the power of the additive noise can often be estimated, for example at a high frequency where the transfer function is zero, and the power of the signal can be estimated at low frequencies. Even if this estimation is not precise and if the noise deviates appreciably from the Gaussian hypothesis, the restoration by using Eq. (\ref{eq:2}) proves  \cite{Wahl1987} to be much better  than a simple multiplication by  $\frac{1}{\alpha}$.

	\subsubsection{Example 2: estimation of the variance of a Gaussian process}
Although all the three above estimators are asymptotically unbiased (i.e. converge to the true value for large $N$), they give very different results for small $N$ in absence of any \textit{a priori} information about $\sigma^2$. Let us consider $N=2$, the minimum number of measurements that gives an information on the variance. We also take $\alpha=1$: we have two measurements $\{d_i=\theta+n_i, i=1, 2\}$ where $n_i$ is an additive independent centered Gaussian noise, of totally unknown variance $\sigma^2$: no \textit{a priori} information is available.  Well known calculations (see Annex 2 for passing from Eq. (\ref{eq:3}) to Eq. (\ref{eq:4})) lead to:
\begin{itemize}
	\item \textbf{Minimum-variance unbiased estimator: }
\begin{equation}
\begin{array}{rcl}
\hat{\sigma}_E^2&=&\displaystyle \frac{1}{N-1}\sum_{i=1}^N(d_i-\bar{d})^2\\
\hat{\sigma}_E^2&=&\displaystyle \frac{1}{2}(d_1-d_2)^2 \quad \textrm{for } N=2
\end{array}\label{eq:3}
\end{equation}\\
	\item \textbf{MMSE estimator: }
\begin{equation}
\begin{array}{rcl}
\hat{\sigma}_M^2&=&\displaystyle \frac{1}{N+1}\sum_{i=1}^N(d_i-\bar{d})^2\\
\hat{\sigma}_M^2&=&\displaystyle \frac{1}{6}(d_1-d_2)^2 \quad \textrm{for } N=2
\end{array}\label{eq:4}
\end{equation}\\
	\item \textbf{\textit{a posteriori}  mean: }
\begin{equation}\\
\begin{array}{rcl}
\me(\sigma^2)&=&\displaystyle \infty \quad \textrm{for } N < 4\\
\me(\sigma^2)&=&\displaystyle \frac{1}{N-3}\sum_{i=1}^N(d_i-\bar{d})^2 \quad \textrm{for } N\geq 4
\end{array}\label{eq:5}
\end{equation}\\
\end{itemize}

The minimum-variance unbiased estimator $\hat{\sigma}_E^2$ given in Eq. (\ref{eq:3})  is known in the time and frequency metrology domain as the Allan variance. It should be certainly used, because of its unbiasedness,  if we can repeat the measure on  many other couples of measurements. However, we restrict our analysis to the case where only $d_1$ and $d_2$ are available, or, at least, where the number of measurements is small.

\subsubsection{Rough explanation of the differences between the estimator results}
Unlike in Example 1, the last two estimators give very different results. An explanation of this difference can be given as follows.

\begin{figure}
(A)\centerline{\includegraphics[height=6.3cm]{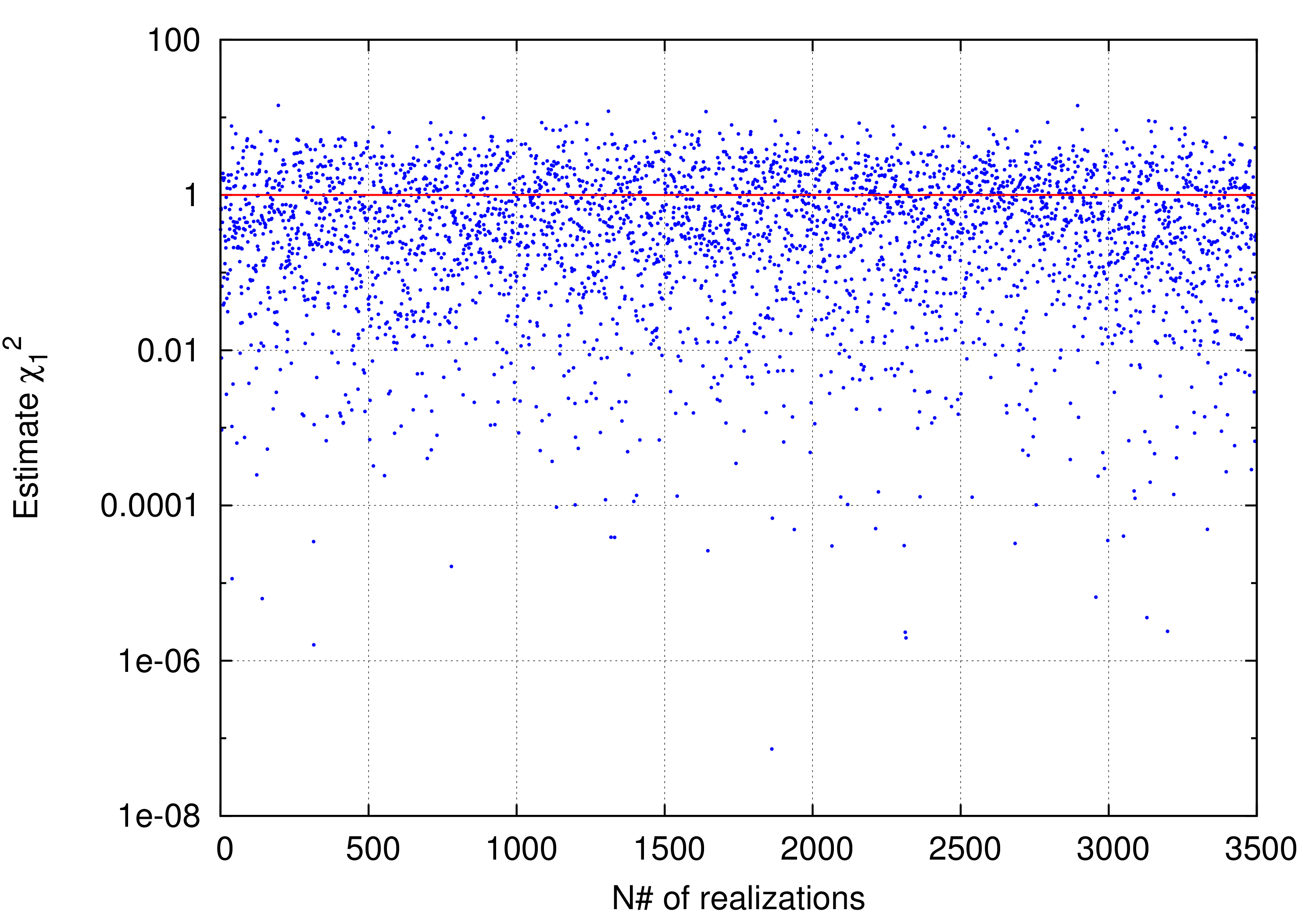}}
(B)\centerline{\includegraphics[height=6.3cm]{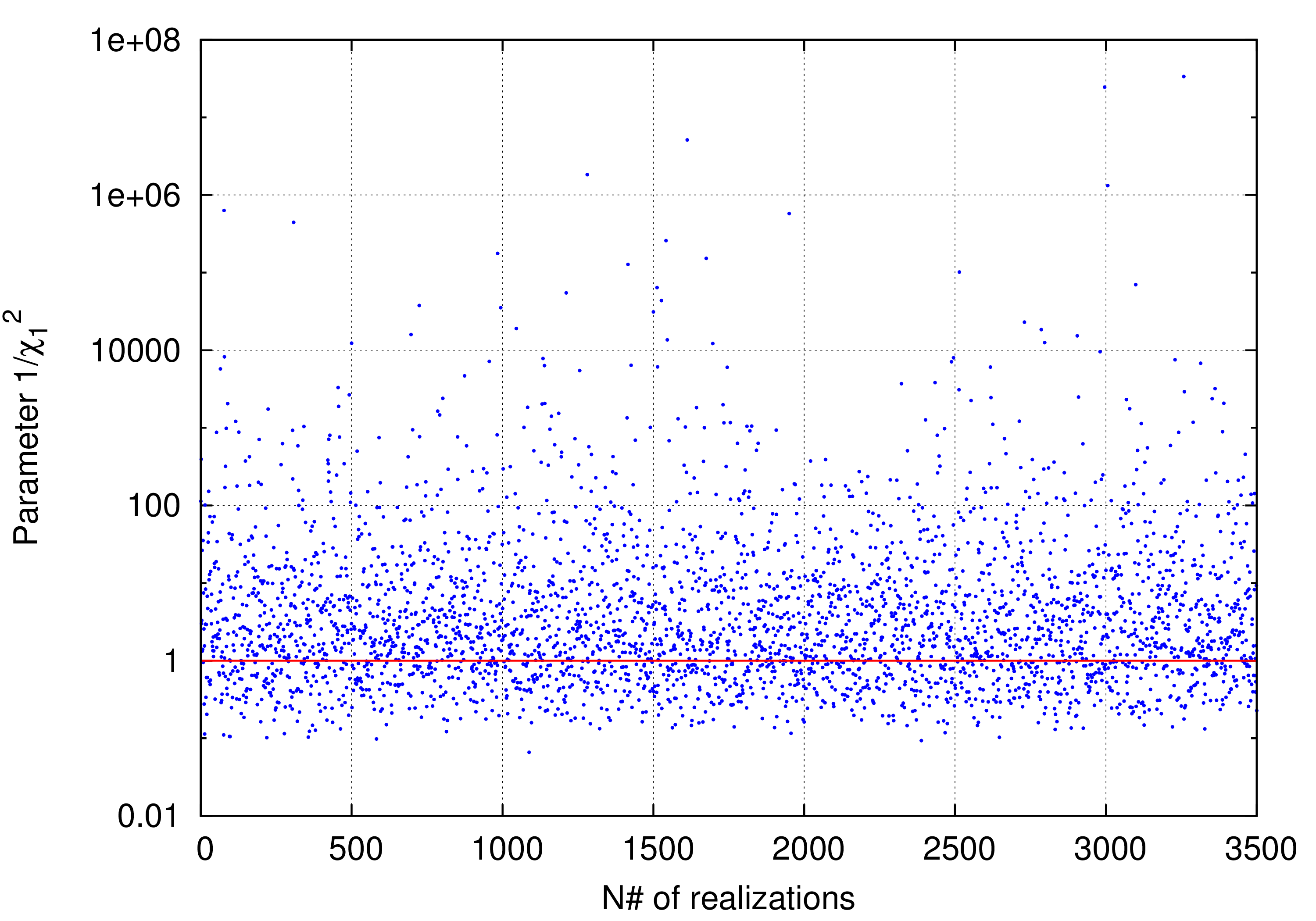}}

\caption{(A): Set of 3500 realizations which follow a $\chi_1^2$ distribution. (B): Set of 3500 realizations which follow an inverse $\chi_1^2$ distribution. 
\label{fig:modele}}
\end{figure}

Let us define, for $N=2$, $Y= \hat{\sigma}_E^2/\sigma^2$. In the model world, $Y $ obeys, as shown in Annex 2 Eq. (\ref{khi}), a $\chi^2$ law with $N-1=1$ degree of freedom (see Figure \ref{fig:modele}(A)), where the estimator $\hat{\sigma}_E^2$ is a random variable and the true value $\sigma^2$ appears as a constant coefficient.

In the measurement world, $\sigma^2$ is a random variable which follows, as shown in Annex 2 Eq. (\ref{1surkhi}), an inverse $\chi_{N-1}^2$ distribution (see Figure \ref{fig:modele}(B)) and $\hat{\sigma}_E^2$ a known constant coefficient issued from the measurements.

  In both worlds, the probability of having a true value $\sigma^2$ much greater than the unbiased estimator $\hat{\sigma}_E^2$ has the same non negligible value : for example $P(\sigma^2>10\ \hat{\sigma}_E^2)=0.25$. However, this probability has completely different consequences in each world.

In the measurement world, the possibly huge values of the true value $\sigma^2$ induce the divergence of the \textit{a posteriori} mean for $N<4$ (huge values of $\frac{1}{\chi_1^2}$ in Eq. (\ref{1surkhi})). These huge values occur in the real world with a non negligible probability and the divergence of the mean is a simple consequence of this existence (see Figure \ref{fig:modele}(B)).

In the model world for a given true value $\sigma^2$, the random realizations of the measurements give, with the same non negligible probability, values such that the estimator $\sigma_E^2$ is much smaller than $\sigma^2$. In other words, the true unknown value of $\sigma^2$ is huge, if expressed in units of $\hat{\sigma}_E^2$, the only available value from the measurements. Unfortunately, these low values of $\hat{\sigma}_E^2$ with respect to the true value have almost no weight in the estimator expectation given by Eq. (\ref{khi}), and also in the MMSE estimator expectation which is proportional to it: less than 1\% of this expectation is due to values of $\hat{\sigma}_E^2<0.1 \sigma^2$, though the probability of having $\hat{\sigma}_E^2<0.1 \sigma^2$ is the same 0.25 as in the measurement world (see Figure \ref{fig:modele}(A)).

\textbf{Because the danger of underestimating the  true value is not properly taken into account, the MMSE estimator is a bad estimator for a scale parameter.} Though not new (see for example \cite{Tarantola1987}), this statement was often missed  \cite{Kay2008}.

\begin{figure}
\includegraphics[width=\columnwidth]{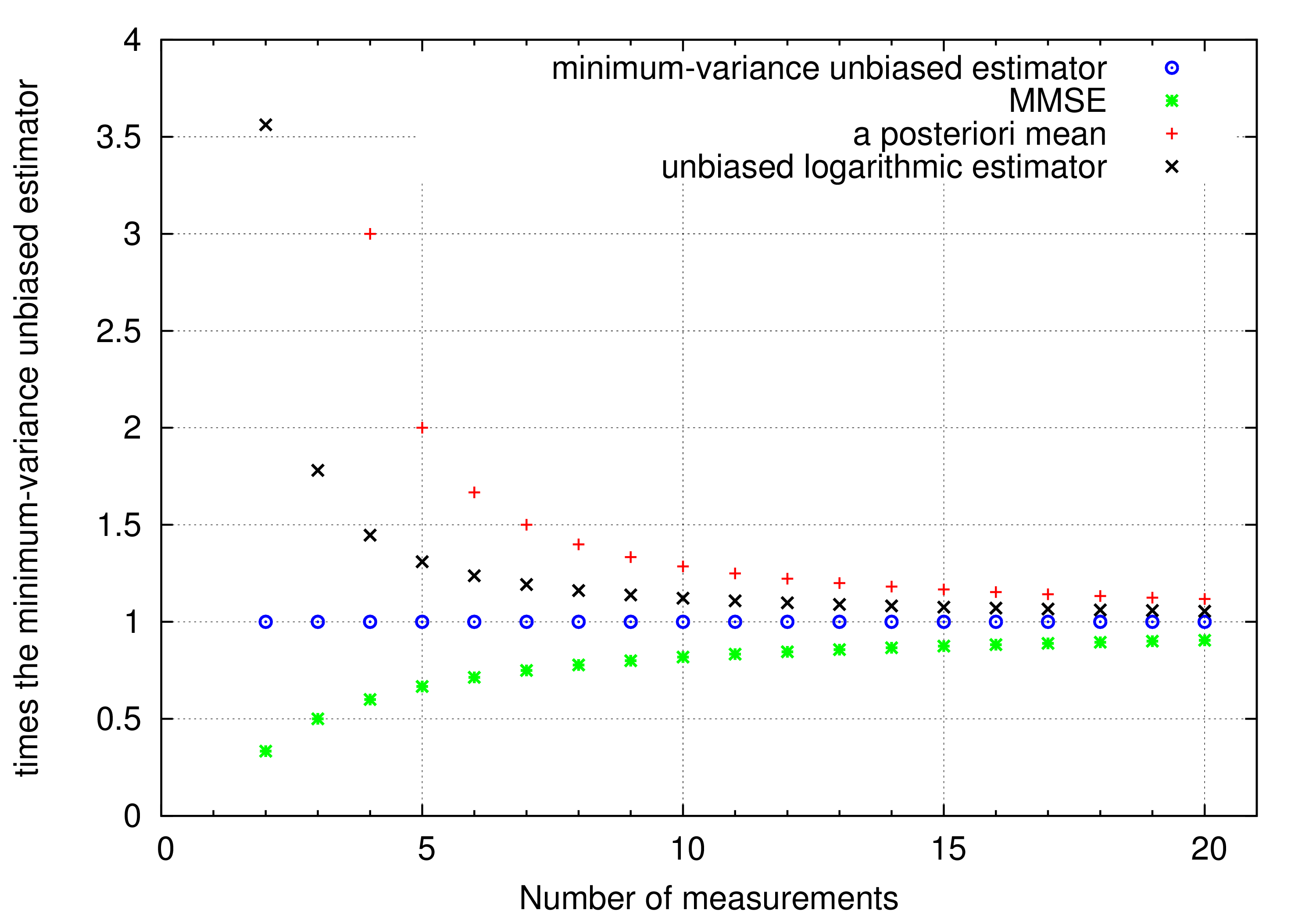}
\caption{Comparison of the three estimators of Section \ref{sec:def}, and of the unbiased logarithmic estimator proposed in Section \ref{sec:opt}.\label{fig:extrap}}
\end{figure}

Even for a greater number of measurements, the difference between the estimators remains non negligible. For instance, the MMSE and \textit{a posteriori} mean differ by  20\% for 20 measurements (see Figure \ref{fig:extrap}).

\section{Solution: An optimal estimator ?\label{sec:opt}}
The situation of Example 2 seems at first sight desperate, since the ``right'' estimator in the measurement world  diverges. The best solution would be, of course, to make more measurements: $\me(\sigma^2)$  is defined for $N \geq 4$. In some cases, this is not possible, especially in the time and frequency metrology domain at very long duration ($10^6-10^7$ s): we would have to wait for several days-months. Moreover, appreciable differences remain between the estimators even for more measurements, as shown in Figure \ref{fig:extrap}. The following considerations give clues to the solution:
\begin{itemize}
	\item A confidence interval on $\sigma^2$ can be defined without any difficulty: at 95 \% of confidence, $\sigma^2$ lies between $0.18$ $\hat{\sigma}_E^2$ and $700$ $\hat{\sigma}_E^2$ since it obeys an inverse $\chi^2$ density with one degree of freedom. The divergence of the mean comes from the values above the high limit of this interval.
	\item Because of this huge confidence interval, only an order of magnitude of $\sigma^2$ can be determined, suggesting that the natural variable choice is $\log(\sigma^2)$.
\item The entire set $\{d_i\}$ can be replaced without any loss of information by an estimator, called a sufficient statistics: the three estimators, Eq. (\ref{eq:3}-\ref{eq:5}), define each a sufficient statistics for the variance of a gaussian distribution, differing only by a known multiplicative constant for a given number of measurements. More generally, $C \cdot \hat{\sigma}_E^2$ is a sufficient statistics whatever the value of the multiplicative constant $C$. Likewise the sample mean is a sufficient statistics for the determination of the mean of a gaussian distribution. \cite{Saporta2011}
	\item To determine the \textit{a posteriori} law $p(\theta)$, we have used the so called ``fiducial argument'', introduced by Fisher \cite{Fisher1973}, which is valid if:
	\begin{enumerate}
		\item no \textit{a priori} information exists, rendering the measurements strictly not recognizable as appertaining to a subpopulation \cite{Fisher1973}
		\item transformations of a sufficient statistics $C\cdot \hat{\sigma}_E^2$  to $u$ and of $\theta$ to $\tau$ exist, such that $\tau$ is a location parameter for the PDF $p(u|\tau)$ \cite{Lindley1958}, i.e. $u=\log(C\cdot \hat{\sigma}_E^2)$ and $\tau=\log(\theta)$, transforming (see Eq. (\ref {scale})) $p_{\theta}(C\cdot \hat{\sigma}_E^2|\theta)=p_{\theta}\left(\frac{C\cdot \hat{\sigma}_E^2}{\theta}\right)$ to $p_{\tau}(u|\tau)=p_{\tau}(u-\tau)$.
\end{enumerate}
After this transformation, the quotients characterizing any scaled probability density, Eq.  (\ref {scale}), become differences and the probability density in both the model and the measurement worlds can be expressed as a function of $u-\tau$: $p(u|\tau)=p(\tau|u )=f(u-\tau)$, implying a constant \textit{a priori} probability density $\pi(\tau)$. Indeed, $p(u\ \textrm{and}\ \tau)=p(\tau|u )p(u)=p(u|\tau)\pi(\tau)$, implying $\pi(\tau)=p(u)$, where $p(u)$ is constant since $u$ is a constant issued from the measurements.
\end{itemize}

If such a transformation exists, the derivation of $p(\theta)$ is warranted: as stated in Eq. (\ref {1surkhi}), $p(\sigma^2|\hat{\sigma}_E^2)$ is an inverse $\chi^2$ density and the expectation of $\sigma^2$ can be calculated. On the other hand, the direct use of $u$  and $\tau$, though not yet the most popular choice, allows a perfect symmetry between both worlds \cite{Vernotte2012}. Indeed, for any scale parameter $\theta$ we have $u=\log(\hat{\theta})+ B$, $\tau=\log(\theta)$, where $B$ is a constant chosen in order to obtain a non biased estimator in the model world:
\begin{equation}
B \quad \textrm{such} \quad \int_{-\infty}^{+\infty}u \cdot p(u|\theta_0) du = \int_{-\infty}^{+\infty} u \cdot f(u-\tau_0) du = \tau_0 \label{eq:6}
\end{equation}\\
where $\theta_0$ is the true value of the parameter $\theta$ and $\tau_0=\log(\theta_0)$.

Then we have in the measurement world, after measurements leading to a given value $u_0$:
\begin{equation}
\int_{-\infty}^{+\infty}\tau \cdot p(\tau|u_0) d\tau = \int_{-\infty}^{+\infty} \tau \cdot f(u_0-\tau) d\tau =u_0. \label{eq:7}
\end{equation}\\
The demonstration is performed by a variable change $x=u-\tau_0$ in Eq. (\ref{eq:6}), leading to $\int_{-\infty}^{+\infty} x f(x)dx=0$ by using $\int_{-\infty}^{+\infty} f(x)dx=1$. Then Eq. (\ref{eq:7})  is obtained by using $y=u_0-\tau$.

In the particular case of the Example 2 with $N=2$, we find  $u=\log(\hat{\sigma}_E^2)+ B$, with $B=1.27$. Hence we proposed \cite{Vernotte2012} in linear units a new estimator $\sigma_L^2=\exp(u)=\exp(B)\hat{\sigma}_E^2=3.56\ \hat{\sigma}_E^2$. This estimator is log-unbiased: $\me\left[\log(\hat{\sigma}_L^2)\right]=\log(\sigma^2)$ and, in  the measurement world $\me\left[\log(\sigma^2)\right]=\log(\hat{\sigma}_L^2)$.  In short, $\log(\hat{\sigma}_L^2)$ is both an unbiased estimator and an \textit{a posteriori} mean.

The above arguments extend to other estimators of a scale parameter. The most evident other example is the estimation of the expected lifetime $\lambda^{-1}$ of a Poisson process, defined as the inverse of the mean rate $\lambda$. Similar considerations allow the definition of a log unbiased estimator, that is, for example for a unique measurement, $1.78$ (i.e. $\exp(e)$, where $e$ is the Euler's constant) times the minimum-variance unbiased estimator. Note that making $N$ measurements consists in waiting from an origin time $t=0$ until the time $t_N$ where the $N^\textrm{th}$ event occurs. The minimum-variance unbiased estimator is $t_N/N$. A wrong procedure would be to define \textit{a priori} a time interval $T$ and to count the number of events in $T$. Such a procedure induces \textit{a priori} information on the magnitude of $\lambda$ and leads to famous absurdities when trying to define unbiased estimators \cite{Romano1986}.

Let us return to Example 1 under the light of the above considerations. The sample mean $\bar{d}$ obeys a Gaussian distribution of mean $\theta$ and variance $\sigma^2/N$ (for $\alpha=1$). Hence, it is directly a location parameter, ensuring that $\bar{d}$ is both a minimum-variance unbiased estimator in the model world and the \textit{a posteriori} mean in the measurement world, \textbf{if no \textit{a priori} information on the mean is available}. This is a great difference with the situation  of Eq. (\ref{eq:2}), where the \textit{a priori} information, i.e. the \textit{a priori} mean power $\me(\theta^2)$ of the signal,  could be taken into account either with the MMSE estimator or with the \textit{a posteriori} mean, but not with an unbiased estimator. Note that, in the measurement world, $p(\bar{d}-\theta)$ is no more Gaussian since $\sigma^2$ is known by its probability density. It is well known that $p\left(\sqrt{N}(\bar{d}-\theta)/\sqrt{\hat{\sigma}_E^2}\right)$ is a Student distribution  for $\bar{d}$ in the model world. Seidenfeld \cite{Seidenfeld1992}, has shown that this law is also, as proposed by Fisher \cite{Fisher1973}, a Bayesian  \textit{a posteriori} law for $\theta$.

Of course, obtaining an unbiased estimator in both worlds is not sufficient to define an optimal estimator. It should also have the minimum variance in the model world. In the measurement world, $p(\theta)$ must be constructed from such a minimum-variance unbiased estimator to ensure a minimum variance on $\theta$. In this case, the estimator will be also  MMSE because the constant \textit{a priori} probability density ensures the same MSE in both worlds for a location parameter. For Example 1 in the absence of any \textit{a priori} information, $p(\theta)$ can be inferred either from the sample average or defined as the mean of the probability density equal to the product of the data likelihoods. Both methods lead to the same results, since the sample mean is a complete sufficient statistics for the underlying Gaussian probability\label{sec:vartheta}.
The minimum-variance unbiased estimator can have more complex forms: for example in the case of a biexponential (or Laplace) distribution of the data, it is obtained by adequate weighting of the ordered data \cite{Govindarajulu1966}. In this case, we have verified  that the mean of the product of the data likelihoods gives the same estimator as the so-called ``efficient estimator'' proposed in \cite{Govindarajulu1966}.

\section{Conclusion}
We have recalled that the three most  popular estimators give very different results for a small number of measurements  in some standard situations. If \textit{a priori} information is available, the difference is irreducible because the best estimator is biased in the direction of this \textit{a priori} information. If no \textit{a priori} information is available, except the model of the underlying probability, these three estimators give the same result for a location parameter. For a scale parameter, using the logarithms of the data allows the transformation of this scale parameter to a location parameter, ensuring the equivalence of the three estimators.

\section{Authors}
\begin{itemize}
	\item \textbf{Eric Lantz} (\texttt{eric.lantz@univ-fcomte.fr}) is professor in the Department of Optics, Femto-ST, at the University of Franche-Comt\'e. 

	\item \textbf{Fran\c{c}ois Vernotte} (\texttt{francois.vernotte@obs-besancon.fr}) is professor in the UTINAM Institute, Observatory THETA, University of Franche-Comt\'{e}.
\end{itemize}

\bibliography{Lantz_Vernotte_paper}

\section*{Annex 1: Wiener filtering}
Let us construct from the data $d_i=\alpha \theta + n_i$ an estimator $\hat{\theta}$:
\begin{equation}
\hat{\theta}=\beta\frac{\bar{d}}{\alpha}=\beta\theta+\frac{\beta}{N\alpha}\sum_{i=1}^Nn_i.
\end{equation}\\
Clearly the real coefficient $\beta$ should approach 1 if the noise term in $\hat{\theta}$ can be neglected, while $\beta$ should approach 0 if this noise becomes predominant. The mean-square error writes:
\begin{equation}
\me\left(\hat{\theta}-\theta\right)^2=(1-\beta)^2\theta^2+\left(\frac{\beta}{N|\alpha|}\right)^2 N\sigma^2,
\end{equation}\\
where we have used our hypotheses on the noise: $n_i$ is centered and additive, i.e. independent of $\theta$, meaning that all cross terms between the  true value and the noise vanish. The determination of the value of $\beta$ minimizing the mean square error is immediate but uses the unknown true value $\theta$:
\begin{equation}
\beta=\frac{N|\alpha|^2\theta^2}{N|\alpha|^2\theta^2+\sigma^2}.
\end{equation}\\
Hence, to use in practice this filter, we have to replace $\theta^2$ by a level of signal $\me(\theta^2)$, known \textit{a priori}, leading to Eq. (\ref{eq:2}). See \cite{Wahl1987} for more details.

\section*{Annex 2: derivation of the MMSE and a posteriori mean for a scale parameter}
\subsection*{Minimum square error}
Following \cite{Kay2008}, we search a coefficient $m$ relating Eq. (\ref{eq:3}) and  (\ref{eq:4}): $m$ is defined by $\hat{\sigma}_M^2=(1-m)\hat{\sigma}_E^2$. Eq. (\ref{eq:1}) becomes, since $\hat{\sigma}_E^2$ is non biased:
\begin{equation}
\me\left[\left(\hat{\sigma}_M^2 - \sigma^2\right)^2\right]=(m\cdot\sigma^2)^2+(1-m)^2\var\left(\hat{\sigma}_E^2\right).
\end{equation}\\
Hence the MSE is minimum for $\frac{\partial}{\partial m} \me\left[\left(\hat{\sigma}_M^2 - \sigma^2\right)^2\right]=0$:
\begin{equation}
m=\frac{\var\left(\hat{\sigma}_E^2\right)}{\var\left(\hat{\sigma}_E^2\right)+\sigma^4}.
\end{equation}\\
Since $(N-1)\frac{\hat{\sigma}_E^2}{\sigma^2}$ obeys a $\chi^2$ distribution with $(N-1)$ degrees of freedom, the variance of $\hat{\sigma}_E^2$ is equal to $\frac{2\sigma^4}{N-1}$, leading to:
\begin{equation}
1-m=\frac{N-1}{N+1}.
\end{equation}
\\
\subsection*{A posteriori mean}
In the model world, where $\sigma^2$ has a definite value, the random variable $\hat{\sigma}_E^2$ can be normalized such that $Y=(N-1)\hat{\sigma}_E^2/\sigma^2$ obeys a $\chi_{N-1}^2$ probability density $p_Y(Y)$, of mean $N-1$. Let $Z=f(Y)$ be a random variable which is a monotonic function of $Y$. If $p_Z(Z)$ is the corresponding probability density, we have $p_Y(Y)dY=p_Z(Z)dZ$, giving immediately for $Z=\hat{\sigma}_E^2$:
\begin{equation}
\me(\hat{\sigma}_E^2)=\int_0^\infty \hat{\sigma}_E^2 p(\hat{\sigma}_E^2|{\sigma}^2)d\hat{\sigma}_E^2 =\frac{\sigma^2}{N-1}\ \me(Y)=\frac{\sigma^2}{N-1}\int_0^\infty  \chi_{N-1}^2 p(\chi_{N-1}^2) d\chi_{N-1}^2=\sigma^2
\label{khi}
\end{equation}
\\
In the measurement world, where $\hat{\sigma}_E^2$ is known, the fiducial argument \cite{Fisher1973} consists in considering the random variable $Z=\sigma^2=(N-1)\hat{\sigma}_E^2/Y$, leading to
\begin{equation}
\me(\sigma^2)=\int_0^\infty\sigma^2p(\sigma^2|\hat{\sigma}_E^2)d\sigma^2=(N-1)\hat{\sigma}_E^2\int_0^\infty\frac{1}{\chi_{N-1}^2}p(\chi_{N-1}^2) d\chi_{N-1}^2. \label{1surkhi}
\end{equation}
\\
This quantity is infinite for $N<4$ and equal to $\frac{(N-1)\hat{\sigma}_E^2}{N-3}$ for $N\geq 4$.  Note that Eq.(\ref{1surkhi}) can be also obtained from a pure Bayesian point of view by introducing an \textit{a priori} law $1/\sigma^2$ to calculate the \textit{a posteriori} law $p(\sigma^2|\hat{\sigma}_E^2)$. Both points of view are equivalent \cite{Lindley1958}.

\end{document}